\documentclass[12pt,reqno,fleqn]{amsart}

\usepackage{latexsym}
\usepackage{amssymb}
\usepackage{amsxtra}
\usepackage{amsmath}
\usepackage{amsfonts}
\usepackage{CJK}
\newcommand{\cleqn}{\setcounter{equation}{0}}
\newcommand{\clth}{\setcounter{theorem}{0}}
\newcommand {\sectionnew}[1]{\section{#1}\cleqn\clth}

\newtheorem{theorem}{Theorem}[section]

\newtheorem{corollary}[theorem]{Corollary}

\renewcommand{\L}{\mathcal{L}}

\newcommand{\R}{\mathbf{R}}

\def\({\left(}
\def\){\right)}
\def\[{\left[}
\def\]{\right]}
\def\d{\partial}

\begin{document}

\title{The recursion operator for a constrained CKP hierarchy}
\author{Chuanzhong Li\dag,  Kelei Tian \dag, Jingsong He$^*$\dag\ddag, Yi Cheng \dag}
\dedicatory { \dag Department of Mathematics, USTC, Hefei, 230026 Anhui, P.\ R.\ China\\
  \ddag Department of Mathematics, NBU, Ningbo, 315211 Zhejiang, P.\ R.\ China}

\thanks{ $^*$ Corresponding author: jshe@ustc.edu.cn, hejingsong@nbu.edu.cn\\
           Supported by:  NSF of China under Grant No. 10671187 and 10971109, and  the Program for  NCET
            under Grant No. NECT-08-0515}

\texttt{}
\date{}

\begin{abstract}
This paper gives a recursion operator for a  1-constrained CKP
hierarchy, and by the recursion operator it  proves that the
1-constrained CKP hierarchy can be reduced  to the mKdV hierarchy
under condition $q=r$.
\end{abstract}

\maketitle
MR(2000) Subject Classification:  37K05, 37K10, 35Q53.\ \ 
\allowdisplaybreaks
 \setcounter{section}{0}

\sectionnew{Introduction}

It is well known that conserved quantities are closely related to
symmetries of equations, and  possessing  infinite number of
conserved quantities or symmetries is a common property of the
classical integrable systems. There are many results on  finding
concrete forms of them \cite{fokos kp , fokos quantities , quantities symmetry}.  Recursion operator is one kind of
effective tools to generate symmetries of the classical integrable
systems\cite{Fuchssteiner Fokas, olver book lie group}. On
the other hand, recursion operator is also used to establish the
Hamiltonian structure of the classical systems
 \cite{fokos kp , fokos symmetry , Strampp Oevel Recursion operators})
  and integrable flows of curves \cite{curves mKdV}.
  So it is vital to
construct the recursion operator for the classical systems. In the
papers \cite{Strampp Oevel Recursion operators, cy1, I, Sokolov recursion operators}, several different
methods are used to construct  recursion operators. Furthermore, it
is highly non-trivial to reduce some results from constrained KP
(cKP) to  constrained BKP  (cBKP) and constrained CKP
 (cCKP) hierarchies which can be seen from
 bilinear forms  \cite{loris Willox bilinear, loris bilinear BKP, I. Loris Bilinear}
 and gauge transformations \cite{He Wu Cheng }. In the paper
\cite{I}, the recursion operator for a 1-constrained cBKP
hierarchy has been given. So the  purpose of this paper is to give
the recursion operator of a cCKP hierarchy and to show the relation
between cCKP hierarchy and mKdV hierarchy.

The organization of this paper is as follows. We recall some basic
facts for the KP hierarchy and a constrained CKP hierarchy in
section 2. In section 3, the recursion operator for the cCKP
hierarchy is discussed and used to generate the $t_3$ flows and
$t_5$ flows, which are consistent with results given by
eigenfunction equations of this sub-hierarchy. Meanwhile we will
show that the $t_3$ flows and $t_5$ flows are the 2-component
generalization of  mKdV equation and 5th order mKdV equation.
Section 4 is devoted to conclusions and discussions in which we will
describe the reducing relations of cKP hierarchy, cBKP hierarchy,
cCKP hierarchy, KdV hierarchy and mKdV hierarchy. 
 \sectionnew{ The constrained CKP hierarchy}

 Since its introduction in 1980s, the KP  hierarchy \cite{dkjm, dl1}
  is one of the most important  research topics  in the
area of classical integrable systems.  The KP  hierarchy is
constructed by the pseudo-differential
operator $L=\d+u_2\d^{-1}+u_3\d^{-2}+....$ like this:\\
$$L_{t_n}=[B_n,L] ,$$ where $B_n=(L^n)_+$. The
$t_2$ (denoted by $y$) flows and $t_3$(denoted by $t$) flows imply
the KP equation
\begin{equation}
(4u_{t}-u_{xxx}-12uu_{x})_x-3u_{yy}=0
\end{equation}
where $u=u_2$. The eigenfunction $q $ and conjugate eigenfunction
$r$ of KP hierarchy are  defined by
\begin{equation} \label{eqeigenfunction}
q_{t_m}=B_m q, \quad r_{t_m}=-B_m^*r.
\end{equation}

 It is well known that there are two kinds
of sub-hierarchies of KP hierarchy, i.e. BKP hierarchy \cite{dkjm}
and  CKP hierarchy \cite{dkjm2}. In order to define the CKP
hierarchy, we need a formal adjoint operation $*$ for an arbitrary
pseudo-differential operator $P=\sum\limits_i p_i \partial^i$,
$P^*=\sum\limits_i (-1)^i\partial^i  p_i$. For example,
$\partial^*=-\partial$, $(\partial^{-1})^*=-\partial^{-1}$, and
$(AB)^*=B^*A^*$ for two operators. The CKP hierarchy is a  reduction
of  the KP hierarchy by the constraint
\begin{equation}\label{CKPconstraint}
L^*=- L ,
\end{equation}
which compresses all even flows of the KP hierarchy, i.e. the Lax
equation of the CKP hierarchy has only odd flows,
\begin{equation}\label{CKPhierarchy}
\dfrac{\partial L}{\partial {t_{2n+1}}}=[B_{2n+1}, L],\ n=0,1,2,
...,
\end{equation}
which indicates  $u_i=u_i(t_1,t_3, t_5,...)$ for the CKP hierarchy.
This hierarchy contains the $(2+1)$ dimensional CKP equation:
\begin{equation}\label{CKP
equation}
9v_{x,t_5}-5v_{t_3t_3}+(v_{xxxxx}+15v_xv_{xxx}+15v^3_x-5v_{xx,t_3}-15v_xv_{t_3}+\frac{45}{4}v^2_{xx})_x=0,
\end{equation}
 where $v=\int{u_2}$. Let $v_{t_3}=0$, eq.(\ref{CKP
equation}) becomes a well-known equation called Kaup-Kupershmidt
equation \cite{KK, KK2} {\allowdisplaybreaks
\begin{equation}
9u_{t_5}+(u_{xxxx}+15uu_{xx}+15u^3+\frac{45}{4}u^2_x)_x=0,
\end{equation}
where $u=u_2$.\\
Moreover, the  so called ``constrained KP hierarchy" (cKP)
\cite{cy1, kss1, cl1})  is a very interesting
sub-hierarchy developed from the point of view of symmetry
constraint, and the Lax operator for 1-constrained KP  is given by}
\begin{equation}\label{cKPLaxoperator}
L=\partial +\sum\limits_{i=1}^n q_i \partial^{-1} r_i,
\end{equation}
here $q_i$ ($r_i$) is the eigenfunction(conjugation eigenfunction)
of $L$ in eq.(\ref{cKPLaxoperator}). By considering CKP condition on
the constrained KP hierarchy, i.e. $L^*=-L$, the constrained CKP
hierarchy (cCKP) can be defined through a following Lax operator
\cite{I2}
\begin{equation}\label{cCKPLaxoperator}
L=\partial +\sum\limits_{i=1}^n (q_i \partial^{-1} r_i+  r_i
\partial^{-1} q_i).
\end{equation}
In the following context, we take $n=1$ for simplicity, i.e.,
\begin{equation}\label{1-cCKPLaxoperator}
\mathcal{L}=\partial + q \partial^{-1} r+  r \partial^{-1} q.
\end{equation}
Note that $q$ and $r$ satisfy the eigenfunction
eqs.(\ref{eqeigenfunction}) associated with $\mathcal{L}$ in
eq.(\ref{1-cCKPLaxoperator}).

 As we know, the evolutions of CKP hierarchy  with respect to
$t_{2},t_{4},t_{6},...$ are freezed. They are also done  to {\bf
1-constrained CKP hierarchy} whose evolution equations are like
this:
\begin{equation}\label{1-CKPhierarchy}
\dfrac{\partial \mathcal{L}}{\partial {t_{2n+1}}}=[B_{2n+1}, \L],\
n=0,1,2, ....
\end{equation}

 In order to get
the explicit form of the flow equations, we need $B_{2n+1}$,
\begin{eqnarray*}
B_1&=&\partial,\\
B_3&=&\d^3+6qr\d+3rq_x+3qr_x,\\
B_5&=&\d^5+10qr\d^3+(15rq_x+15qr_x)\d^2+(15qr_{xx}+15rq_{xx}+40q^2r^2+20q_xr_x)\d\\
&&+40qr^2q_x+40rq^2r_x+5qr_{xxx}+5rq_{xxx}+10q_xr_{xx}+10r_xq_{xx}.\\
\end{eqnarray*}
After a direct computation from eigenfunction
eqs.(\ref{eqeigenfunction}), we can get  the first few flows of the
cCKP hierarchy
\begin{equation}\label{t1flow}
\begin{cases}
q_{t_1}=q_x\\
r_{t_1}=r_x,\\
\end{cases}
\end{equation}
\begin{equation}\label{t3flow}
\begin{cases}
q_{t_3}=q_{xxx}+9qrq_x+3q^2r_x\\
r_{t_3}=r_{xxx}+9qrr_x+3r^2q_x,\\
\end{cases}
\end{equation}
\begin{equation}\label{t5flow}
\begin{cases}
q_{t_5}&=q_{xxxxx}+15qrq_{xxx}+30rq_xq_{xx}+25qr_xq_{xx}+25qq_xr_{xx}\\
&+80q^2r^2q_x+20q_xr_xq_x+40rq^3r_x+5q^2r_{xxx}\\
r_{t_5}&=r_{xxxxx}+15qrr_{xxx}+30qr_xr_{xx}+25rq_xr_{xx}+25rr_xq_{xx}\\
&+80q^2r^2r_x+20q_xr_xr_x+40qr^3q_x+5r^2q_{xxx}.\\
\end{cases}
\end{equation}
Let $q=r$, eq.(\ref{t3flow}) implies mKdV equation
\begin{equation}\label{t3flow 2}
q_{t_3}=q_{xxx}+12q^2q_x.
\end{equation}
A transformation $q=\frac{\sqrt{3}u}{6}$ leads it to the form of
mKdV equation in (\cite{Olver mkdv})
\begin{equation}\label{t3flow 2}
u_{t_3}=u_{xxx}+u^2u_x.
\end{equation}

Let $q=r$, eq.(\ref{t5flow}) implies 5th order mKdV equation
 \begin{equation}\label{t5flow 2}
q_{t_5}=q_{xxxxx}+20q^2q_{xxx}+80qq_xq_{xx}+120q^4q_x+20(q_x)^3.
\end{equation}
A transformation $q=\frac{\sqrt{3}u}{6}, t_5=t$ leads it to the
standard 5th order mKdV equation in \cite{Olver mkdv}
 \begin{equation}\label{t5flow 3}
u_t=u_{xxxxx}+\frac{5}{3}u^2u_{xxx}+\frac{20}{3}uu_xu_{xx}+\frac{5}{6}u^4u_x+\frac{5}{3}(u_x)^3.
\end{equation}
Note that there exist other forms of mKdV equation and 5th order
mKdV equation, for example \cite{I2, M. Ito}.
 It is very difficult to observe recursion operator from
equations on $t_3$ flows and $t_5$ flows above.  We shall find it in
next section from eigenfunction equations on $q$ and $r$, and may
use it to generate any higher order flows. To illustrate the
validity of recursion operator, we shall use it to generate $t_3$
flows from  trivial flows, i.e. $t_1$ flows, and further generate
$t_5$ flows from $t_3$ flows.

\sectionnew{Recursion operator}
In this section, we will give the form of recursion operator $R$.
Now, we define the following four operators:
\begin{eqnarray*}
R_{11}&=&\L^2+3qr+\L(r)\d^{-1}q +2q_x\d^{-1}r-q\d^{-1}qr\d^{-1}r-r\d^{-1}q\d-r\d^{-1}q^2\d^{-1}r\\
&&-2r\d^{-1}rq\d^{-1}q-r\d^{-1}q(\int{rq})-q\d^{-1}q(\int{r^2}),\\
R_{12}&=&2q_x\d^{-1}q+3q^2-2q\d^{-1}q^2\d^{-1}r-q\d^{-1}q\d-q\d^{-1}qr\d^{-1}q-r\d^{-1}q^2\d^{-1}q\\
&&-q\d^{-1}q(\int{rq})-r\d^{-1}q(\int{q^2})+\L(q)\d^{-1}q,\\
R_{21}&=&2r_x\d^{-1}r+3r^2-2r\d^{-1}r^2\d^{-1}q-r\d^{-1}r\d-r\d^{-1}qr\d^{-1} r-q\d^{-1}r^2\d^{-1}r\\
&&-r\d^{-1}r(\int{rq})-q\d^{-1}r(\int{r^2})+\L(r)\d^{-1}r,\\
R_{22}&=&\L^2+3qr+\L(q)\d^{-1}r+2r_x\d^{-1}q-r\d^{-1}qr\d^{-1}q-q\d^{-1}r\d-q\d^{-1}r^2\d^{-1}q\\
&& -2q\d^{-1}qr\d^{-1} r-q\d^{-1}r(\int{rq})-r\d^{-1}r(\int{q^2}).
\end{eqnarray*}
\begin{theorem}\label{theorem}
The recursion relation of flows for the 1-cCKP hierarchy
(\ref{1-CKPhierarchy}) is like this:
\begin{equation}
\label{recursion operator} \left( {\begin{array}{*{20}c}
   {{q}}  \\
   {{r}}  \\
\end{array}} \right)_{{\rm{t}}_{{\rm{m + 2}}} }  = \left( {\begin{array}{*{20}c}
   {R_{11} } & {R_{12} }  \\
   {R_{21} } & {R_{22} }  \\
\end{array}} \right) \left( {\begin{array}{*{20}c}
   {{q}}  \\
   {{r}}  \\
\end{array}} \right)_{{\rm{t}}_{\rm{m}}}.
\end{equation}
\end{theorem}
\textbf{Proof}.
 Using the identities (\cite{Aratyn }) below
\begin{eqnarray}\label{identity'}
(B_nf\d^{-1}g)_{-}&=&B_n(f)\d^{-1}g, \\
\label{identity}
 (f\d^{-1}gB_n)_{-}&=&f\d^{-1}B_n^*(g),\\
 f_1\d^{-1}g_1\cdot f_2\d^{-1}g_2&=&f_1(\int {g_1f_2})\d^{-1}g_2-f_1\d^{-1}g_2(\int
 {f_2g_1}),
\end{eqnarray}
 we can calculate the $\L^2$ as following:
\begin{equation}
\L^2=\d^2+4qr+q\d^{-1}\L^*(r)+r\d^{-1}\L^*(q)+\L(q)\d^{-1}r+\L(r)\d^{-1}q
\end{equation}
with
$$\L(q)=q_x+q\int{rq}+r\int{q^2} ,    \quad \L(r)=r_x+q\int{r^2}+r\int{qr},$$
$$\L^*(q)=-q_x-q\int{rq}-r\int{q^2} ,    \quad \L^*(r)=-r_x-q\int{r^2}-r\int{qr}.$$
Therefore,
$$B_2=\d^2+4qr.$$
Denote $A_n$ as $(\L^n)_-$, $n=1,2,....$
 Considering CKP condition and eqs.(\ref{eqeigenfunction}), $q$ and $r$ should
 satisfy the same equation, i.e.
\begin{equation}\label{eigenfunction qr}
\ \ \ \ B_m(q)=q_{t_{m}}, \ \ \ B_m(r)=r_{t_{m}},
\end{equation} then
\begin{eqnarray}\label{recursion relation}
q_{t_{m+2}}=(\L^2\L^m)_+q=B_2B_mq+(B_2A_m)_+q+(A_2B_m)_+q,
\end{eqnarray}
\begin{eqnarray}\label{recursion relation r}
r_{t_{m+2}}=(\L^2\L^m)_+r=B_2B_mr+(B_2A_m)_+r+(A_2B_m)_+r.
\end{eqnarray}

Firstly, we will calculate  $(B_2A_m)_+$. Now, we set
$A_m=\d^{-1}a_1+\d^{-2}a_2+...$.\\
So, $(B_2A_m)_+=\d a _1+a_2$. The identity $Res_{\d}[\L^m,\L]=0$
yields:
\begin{equation}\label{res1}
Res_{\d}[B_m, \L]=Res_{\d}[-A_m, \L]=Res_{\d}[-A_m, B_1].
\end{equation}
The first residue  of eq.(\ref{res1}) equals
$Res_{\d}\L_{t_m}=2(qr)_{t_m}$, the last residue  of eq.(\ref{res1})
yields $Res_{\d}[\d, \d^{-1}a_1+\d^{-2}a_2+...]=a_{1x}$.
So,\begin{equation}a_1=\int{2(qr)_{t_m}}.\end{equation}

To compute $a_2$, we should use identity $Res_{\d}[\L^m,\L^2]=0$,
considering the similar identity
\begin{equation}\label{res2}
Res_{\d}[B_m,\L^2]=Res_{\d}[-A_m,\L^2]=Res_{\d}[-A_m,B_2].
\end{equation}
The first residue of eq.(\ref{res2}) equals $Res_{\d}\L^2_{t_m}=0$,
the last residue  of eq.(\ref{res2}) yields $Res_{\d}[\d^2+4qr,
\d^{-1}a_1+\d^{-2}a_2+...]=-a_{1xx}+2a_{2x}$. We can easily get
\begin{equation}a_2=a_{1x}/2=(qr)_{t_m}.
\end{equation}
 Hence,\begin{equation}(B_2A_m)_+=\d\cdot
\int{2(qr)_{t_m}}+(qr)_{t_m}.\end{equation}
 About the term
$(A_2B_m)_+$, we write it as $A_2B_m-(A_2B_m)_-$. The first term is
relevant to $t_m$ flow. Using the identity\eqref{identity}, we can
compute  the second term
\begin{eqnarray*}
(A_2B_m)_-&=&[(q\d^{-1}\L^*(r)+r\d^{-1}\L^*(q)+\L(q)\d^{-1}r+\L(r)\d^{-1}q)B_m]_-\\
&=&q\d^{-1}B_m^*\L^*(r)+r\d^{-1}B_m^*\L^*(q)+\L(q)\d^{-1}B_m^*(r)+\L(r)\d^{-1}B_m^*(q).
\end{eqnarray*}
 Considering  eqs.(\ref{eigenfunction qr}),
\begin{eqnarray*}
B_m^*\L^*(q)&=&\L^*B_m^*(q)+[B_m^*,\L^*](q)\\
&=&\L B_m(q)+[B_m,\L](q)\\
&=&\L(q_{t_{m}})+\L_{t_{m}}(q)\\
&=&q_{xt_m}+r\int qq_{t_m}+q\int
rq_{t_m}\\
&&+(r_{t_m}\d^{-1}q+r\d^{-1}q_{t_m}+q_{t_m}\d^{-1}r+q\d^{-1}r_{t_m})(q)\\
&=&q_{xt_m}+2r\int qq_{t_m}+q\int rq_{t_m}+r_{t_m}\int
q^2+q_{t_m}\int rq+q\int r_{t_m}q.
\end{eqnarray*}

Similarly, we can get
\begin{eqnarray*}
B_m^*\L^*(r) &=&r_{xt_m}+2q\int rr_{t_m}+r\int qr_{t_m}+q_{t_m}\int
r^2+r_{t_m}\int qr+r\int q_{t_m}r.
\end{eqnarray*}

After bringing these results into eq.\eqref{recursion relation}, we
get the recursion flow of  $q$
\begin{eqnarray*}
q_{t_{m+2}} &=&\Big[\L^2+3qr+\L(r)\d^{-1}q
+2q_x\d^{-1}r-q\d^{-1}qr\d^{-1}r-r\d^{-1}q\d-r\d^{-1}q^2\d^{-1}r\\
&&-2r\d^{-1}rq\d^{-1}q
-r\d^{-1}q(\int{rq})-q\d^{-1}q(\int{r^2})\Big]q_{t_m}\\
&&+\Big[2q_x\d^{-1}q+3q^2-2q\d^{-1}q^2\d^{-1}r-q\d^{-1}q\d-q\d^{-1}qr\d^{-1}q-r\d^{-1}q^2\d^{-1}q\\
&&
-q\d^{-1}q(\int{rq})-r\d^{-1}q(\int{q^2})+\L(q)\d^{-1}q\Big]r_{t_m}.\\
\end{eqnarray*}

Similarly after bringing these results into eq.\eqref{recursion
relation r}, we get the recursion flow of  $r$

\begin{eqnarray*}
r_{t_{m+2}}&=&
\Big[2r_x\d^{-1}r+3r^2-2r\d^{-1}r^2\d^{-1}q-r\d^{-1}r\d-r\d^{-1}qr\d^{-1} r-q\d^{-1}r^2\d^{-1}r\\
&&-r\d^{-1}r(\int{rq})-q\d^{-1}r(\int{r^2})+\L(r)\d^{-1}r\Big]q_{t_m}\\
&&+\Big[\L^2+3qr+\L(q)\d^{-1}r+2r_x\d^{-1}q-r\d^{-1}qr\d^{-1}q-q\d^{-1}r\d-q\d^{-1}r^2\d^{-1}q \\
&&-2q\d^{-1}qr\d^{-1} r -q\d^{-1}r(\int{rq})-r\d^{-1}r(\int{q^2})\Big]r_{t_m}.\\
\end{eqnarray*}
Then we get the recursion operator  written in eq.(\ref{recursion
operator}).   \qed\\
Now, let us inspect whether  the results from this recursion
operator are consistent with what from the eigenfunction eqs.(\ref{eqeigenfunction}).\\
By a very tedious calculation, we have checked that they are consistent on the $t_3$ flows
and $t_5$ flows. Of course we can generate the $t_7$ flows, $t_9$
flows etc. in the same way which should be also consistent with the
corresponding flows from Sato's methods.
\begin{corollary}\label{corollary}
The 1-constrained CKP hierarchy (\ref{1-CKPhierarchy}) can be
reduced  to the mKdV hierarchy  by condition $q=r$.
\end{corollary}
 \textbf{Proof}. Let $q=r$, we can get
\begin{eqnarray*}
q_{t_{m+2}} &=&\Big[\L^2(q,q)+3qq+\L(q)\d^{-1}q
+2q_x\d^{-1}q-q\d^{-1}qq\d^{-1}q-q\d^{-1}q\d-q\d^{-1}q^2\d^{-1}q\\
&&-2q\d^{-1}qq\d^{-1}q
-q\d^{-1}q(\int{qq})-q\d^{-1}q(\int{q^2})\Big]q_{t_m}\\
&&+\Big[2q_x\d^{-1}q+3q^2-2q\d^{-1}q^2\d^{-1}q-q\d^{-1}q\d-q\d^{-1}qq\d^{-1}q-q\d^{-1}q^2\d^{-1}q\\
&&
-q\d^{-1}q(\int{qq})-q\d^{-1}q(\int{q^2})+\L(q)\d^{-1}q\Big]q_{t_m}\\
 &=&\Big[\d^2+4qq+q\d^{-1}(-q_x-q\int{q^2}-q\int{qq})+q\d^{-1}(-q_x-q\int{qq}-q\int{q^2})\\
&&+2(q_x+q\int{qq}+q\int{q^2})\d^{-1}q+(q_x+q\int{q^2}+q\int{qq})\d^{-1}q+3qq\\
&&+2q_x\d^{-1}q-q\d^{-1}qq\d^{-1}q-q\d^{-1}q\d-q\d^{-1}q^2\d^{-1}q\\
&&-2q\d^{-1}qq\d^{-1}q
-q\d^{-1}q(\int{qq})-q\d^{-1}q(\int{q^2})\Big]q_{t_m}\\
&&+\Big[2q_x\d^{-1}q+3q^2-2q\d^{-1}q^2\d^{-1}q-q\d^{-1}q\d-q\d^{-1}qq\d^{-1}q-q\d^{-1}q^2\d^{-1}q\\
&&
-q\d^{-1}q(\int{qq})-q\d^{-1}q(\int{q^2})+(q_x+q\int{qq}+q\int{q^2})\d^{-1}q\Big]q_{t_m}\\
 &=&\Big[\d^2+10q^2+q\d^{-1}(-2q_x-8q\int{q^2}-8qq\d^{-1}q-2q\d)+(8q_x\\
 &&+8q\int{qq})\d^{-1}q\Big]q_{t_m}\\
 &=&(\d^2+8q^2+8q_x\d^{-1}q)q_{t_m}.\\
\end{eqnarray*}
Then we can get the reduced recursion operator which is just the
recursion operator for mKdV hierarchy
\begin{eqnarray}
 R_r&=&\d^2+8q^2+8q_x\d^{-1}q.
\end{eqnarray}
The same transformation $q=\frac{\sqrt{3}u}{6}$ leads to the form of
mKdV hierarchy in \cite{Olver mkdv}
\begin{eqnarray}
 \R&=&\d^2+\frac{2}{3}u^2+\frac{2}{3}u_x\d^{-1}u.
\end{eqnarray}
So we can get the whole mKdV hierarchy from the trivial flow under
the condition $q=r$. For example, mKdV eq.(\ref{t3flow 2}) and  5th
order mKdV eq.\ref{t5flow 2} can be got from this cCKP hierarchy.
Now we will say that condition $q=r$ can reduce the cCKP hierarchy
to mKdV hierarchy. \qed

\sectionnew{Conclusions and Discussions}
The recursion operator in eq.(\ref{recursion operator}) for a cCKP
system was found from the eigenfunction equations on $q$ and $r$.
This operator was used to generate $t_3$ flows (eqs.(\ref{t3flow}))
and $t_5$ flows (eqs.(\ref{t5flow})) from the $t_1$ flows of this
special hierarchy, which are consistent with flows got from
eigenfunction eqs.(\ref{eigenfunction qr}). That demonstrated the
validity of the recursion operator. Of course one can also use it to
generate higher order flows. On the other hand, our results are more
complicated than recursion operator for cKP hierarchy \cite{cy1}.
Moreover, we can also get the following reduction chain from
corollary \ref{corollary}:
\begin{eqnarray}
  \label{reuction digram cCKP}
  {\text{cKP hierarchy} } \xrightarrow{{L^* = - L}}{\text{cCKP hierarchy} } \xrightarrow{{{q} = {{r}}}}
  {\text{m}}{\text{KdV hierarchy} } \hfill.
\end{eqnarray}

Similarly, the KdV hierarchy will appear in the reduction of cBKP
hierarchy \cite{I}. As we know, the relationship of KdV hierarchy
and mKdV hierarchy can be represented by miura transformation, but
what is the similar transformation between cBKP hierarchy and cCKP
hierarchy. In \cite{Adler}, the relationship of KdV hierarchy and
mKdV hierarchy can be seen from the decomposition of differential
Lax operator, but whether the relationship of cBKP hierarchy and
cCKP hierarchy can be comprehended from the the decomposition of
pseudo-differential Lax operator is still unknown and interesting.

 {\bf {Acknowledgments:}}\\
  {\small
   We thank Professor Li Yishen (USTC, China) for long-term encouragements and supports. }

\end{document}